\documentstyle[11pt]{article}
\input amssymb.sty

\title{Modality, Potentiality and Contradiction\\in Quantum Mechanics}

\author{{\sc Christian de Ronde}}
\date{}

\begin{document}

\bibliographystyle{plain}
\maketitle

\begin{center}
\begin{small}
Instituto de Filosof\'ia ``Dr. A. Korn" \\ 
Universidad de Buenos Aires, CONICET - Argentina \\
Center Leo Apostel and Foundations of  the Exact Sciences\\
Brussels Free University - Belgium \\
\end{small}
\end{center}

\begin{abstract}
\noindent In \cite{daCostadeRonde13}, Newton da Costa together with the author of this paper argued in favor of the possibility to consider quantum superpositions in terms of a paraconsistent approach. We claimed that, even though most interpretations of quantum mechanics (QM) attempt to escape contradictions, there are many hints that indicate it could be worth while to engage in a research of this kind. Recently, Arenhart and Krause \cite{ArenhartKrause14a, ArenhartKrause14b, ArenhartKrause14c} have raised several arguments against this approach and claimed that ---taking into account the square of opposition--- quantum superpositions are better understood in terms of {\it contrariety propositions} rather than {\it contradictory propositions}. In \cite{deRonde14a} we defended the Paraconsistent Approach to Quantum Superpositions (PAQS) and provided arguments in favor of its development. In the present paper we attempt to analyze the meanings of {\it modality}, {\it potentiality} and {\it contradiction} in QM, and provide further arguments of why the PAQS is better suited, than the Contrariety Approach to Quantum Superpositions (CAQS) proposed by Arenhart and Krause, to face the interpretational questions that quantum technology is forcing us to consider.
\end{abstract}
\begin{small}


\end{small}

\bibliography{pom}

\begin{thebibliography}{10}


\bibitem{ArenhartKrause14a} Arenhart, J. R. and Krause, D., 2014, ``Oppositions in Quantum Mechanics",  in {\it New dimensions of the square of opposition}, Jean-Yves B\'eziau and Katarzyna Gan-Krzywoszynska (Eds.), 337-356, Philosophia Verlag, Munich.

\bibitem{ArenhartKrause14b} Arenhart, J. R. and Krause, D., 2014, ``Contradiction, Quantum Mechanics, and the Square of Opposition", {\it Logique et Analyse}, accepted. 

\bibitem{ArenhartKrause14c} Arenhart, J. R. and Krause, D., 2014, ``Potentiality and Contradiction in Quantum Mechanics.", in {\it The Road to Universal Logic (volume II)}, Arnold Koslow and Arthur Buchsbaum (Eds.), Springer, in press. 

\bibitem{Bacciagaluppi96} Bacciagaluppi, G., 1996, {\it Topics
in the Modal Interpretation of Quantum Mechanics}, Doctoral
dissertation, University of Cambridge, Cambridge.

\bibitem{Nature13} Bernien, H., Hensen, B., Pfaff, W., Koolstra, G., Blok, M. S., Robledo, L., Taminiau, T. H., Markham, M., Twitchen, D. J., Childress, L. and Hanson, R., 2013, ``Heralded entanglement between solid-state qubits separated by three metres'',  {\it Nature}, {\bf 497}, 86-90.

\bibitem{Beziau12}  B\'eziau, J.-Y., 2012, ``The Power of the Hexagon'', {\it Logica Universalis}, {\bf 6}, 1-43.

\bibitem{Beziau14}  B\'eziau, J.-Y., 2014, ``Paraconsistent logic and contradictory viewpoint'', to appear in {\it Revista Brasileira de Filosofia}, 241.

\bibitem{BokulichBokulich} Bokulich, P., and Bokulich, A., 2005, ``Niels Bohr's Generalization of Classical Mechanics'', {\it Foundations of Physics}, {\bf 35}, 347-371.

\bibitem{Bub97} Bub, J., 1997, {\it Interpreting the Quantum World}, Cambridge University Press, Cambridge.

\bibitem{Nature11a} Clausen, C., Usmani, I., Bussi\`eres, F., Sangouard, N., Afzelius, M., de Riedmatten, H. and Gisin, N., 2011, ``Quantum storage of photonic entanglement in a crystal'',  {\it Nature}, {\bf 469}, 508-511.

\bibitem{daCostadeRonde13} da Costa, N. and de Ronde, C., 2013, ``The Paraconsistent Logic of Quantum Superpositions", {\it Foundations of Physics}, {\bf 43}, 845-858.

\bibitem{daCostadeRonde15} da Costa, N. and de Ronde, C., 2015, ``The Paraconsistent Approach to Quantum Superpositions Reloaded: Formalizing Contradictory Powers in the Potential Realm", in preparation.

\bibitem{deRonde10} de Ronde, C., 2010, ``For and Against Metaphysics in the Modal Interpretation of Quantum Mechanics", {\it Philosophica}, {\bf 83}, 85-117.

\bibitem{deRonde11} de Ronde, C., 2011, {\it The Contextual and Modal Character of Quantum
Mechanics: A Formal and Philosophical Analysis in the Foundations of Physics}, PhD dissertation, Utrecht University.

\bibitem{deRonde13a} de Ronde, C., 2013, ``Quantum Superpositions and Causality: On the Multiple Paths to the Measurement Result", {\it Los Alamos Archive}, arXiv:1310.4534.

\bibitem{deRonde13b} de Ronde, C., 2013, ``Representing Quantum Superpositions: Powers, Potentia and Potential Effectuations", {\it Los Alamos Archive}, arXiv:1312.7322.

\bibitem{deRonde14a} de Ronde, C., 2014, ``A Defense of the Paraconsistent Approach to Quantum Superpositions (Answer to Arenhart and Krause)", {\it Logique et Analyse}, sent. 

\bibitem{deRonde15a} de Ronde, C., 2015, ``Hilbert space quantum mechanics is contextual. (Reply to R. B. Griffiths)", {\it Studies in History and Philosophy of Modern Physics}, sent.

\bibitem{deRonde15b} de Ronde, C., 2015, ``Quantum Superpositions Do Exist! But `Quantum Physical Reality $\neq$ Actuality' (Reply to Dieks and Griffiths)", preprint.

\bibitem{RFD14a} de Ronde, C., Freytes, H. and Domenech, G., 2014, ``Interpreting the Modal Kochen-Specker Theorem: Possibility and Many Worlds in Quantum Mechanics'', {\it Studies in History and Philosophy of Modern Physics}, {\bf 45}, pp. 11-18.

\bibitem{RFD14b}  de Ronde, C., Freytes, H. and Domenech, G., 2014, ``Quantum Mechanics and the Interpretation of the Orthomodular Square of Opposition'', in {\it New dimensions of the square of opposition}, Jean-Yves B\'eziau and Katarzyna Gan-Krzywoszynska (Eds.), 223-242, Philosophia Verlag, Munich.

\bibitem{deRondeMassri14} de Ronde, C. and Massri, C., 2014, ``Revisiting the Orthodox Interpretation of `Physical States' in Quantum Mechanics",  {\it Los Alamos Archive}, arXiv:1412.2701.

\bibitem{Dieks89a} Dieks, D., 1989, ``Quantum Mechanics Without
the Projection Postulate and Its Realistic Interpretation'', {\it
Foundations of Physics}, {\bf 19}, 1397-1423.

\bibitem{Dieks10} Dieks, D., 2010, ``Quantum Mechanics, Chance and Modality", {\it Philosophica}, {\bf 83}, 117-137.

\bibitem{Dirac74} Dirac, P. A. M., 1974, {\it The Principles of
Quantum Mechanics}, 4th Edition, Oxford University Press, London.

\bibitem{DFR06} Domenech, G., Freytes, H. and de Ronde, C., 2006, ``Scopes and limits of modality in quantum mechanics", \textit{Annalen der Physik}, {\bf 15}, 853-860.

\bibitem{DFR08a} Domenech, G., Freytes, H. and de Ronde, C., 2008, ``A topological study of contextuality and modality in quantum mechanics'', {\it International Journal of Theoretical Physics},
{\bf  47}, 168-174.

\bibitem{DFR08b} Domenech, G., Freytes, H. and de Ronde, C., 2009, ``Modal-type orthomodular logic'', {\it Mathematical Logic Quarterly}, {\bf 3}, 307-319.

\bibitem{DFR09} Domenech, G., Freytes, H. and de Ronde, C., 2009, ``Many worlds and modality in the interpretation of quantum mechanics: an algebraic approach'', {\it Journal of Mathematical
Physics}, {\bf 50}, 072108.

\bibitem{FRD12}  Freytes, H., de Ronde, C. and Domenech, G., 2012, ``The square of opposition in orthodmodular logic'', in {\it Around and Beyond the Square of Opposition: Studies in Universal Logic}, Jean-Yves B\'eziau and Dale Jacquette (Eds.), pp. 193-201, Springer, Basel.

\bibitem{FuchsPeres00} Fuchs, C. and Peres, A., 2000, ``Quantum theory needs no `interpretation' '', {\it Physics Today}, {\bf 53}, 70.

\bibitem{Griffiths13} Griffiths, R. B., 2013, ``Hilbert space quantum mechanics is non contextual",  {\it Studies in History and Philosophy of Modern Physics}, {\bf 44}, 174-181.

\bibitem{Heis58} Heisenberg, W., 1958, {\it Physics and Philosophy},
World perspectives, George Allen and Unwin Ltd., London.

\bibitem{NaturePhy12} Ma, X., Zotter, S., Kofler, J., Ursin, R., Jennewein, T., Brukner, C. and Zeilinger, A.,  2012, ``Experimental delayed-choice entanglement swapping'',  {\it Nature Physics}, {\bf 8}, 480-485.

\bibitem{Nature07} Ourjoumtsev, A., Jeong, H., Tualle-Brouri, R. and
Grangier, P., 2007, ``Generation of optical `Schr\"{o}dinger cats'
from photon number states'', {\it Nature}, {\bf 448}, 784-786.

\bibitem{Priest87} Priest, G., 1987, {\it In Contradiction}. Nijhoff, Dordrecht.

\bibitem{Redei} R\'edei, M., 2012, ``Some Historical and Philosophical Aspects of Quantum Probability Theory and its Interpretation'', in {\it Probabilities, Laws, and Structures}, D. Dieks et al. (Eds.), pp. 497-506, Springer, 

\bibitem{Schr35} Schr\"odinger, E., 1935, ``The Present Situation
in Quantum Mechanics'', {\it Naturwiss}, {\bf 23}, 807. Translated
to english in {\it Quantum Theory and Measurement}, J. A. Wheeler
and W. H. Zurek (Eds.), 1983, Princeton University Press, Princeton.

\bibitem{Smets05} Smets, S., 2005,  ``The Modes of Physical Properties
in the Logical Foundations of Physics'', {\it Logic and Logical
Philosophy}, {\bf 14}, 37-53.

\bibitem{VF91} Van Fraassen, B. C., 1991, {\it Quantum Mechanics: An
Empiricist View}, Clarendon, Oxford.

\bibitem{VerelstCoecke} Verelst, K. and Coecke, B., 1999, ``Early
Greek Thought and perspectives for the Interpretation of Quantum
Mechanics: Preliminaries to an Ontological Approach'', In {\it The
Blue Book of Einstein Meets Magritte}, 163-196, D. Aerts (Ed.),
Kluwer Academic Publishers, Dordrecht.

\bibitem{WZ} Wheeler, J. A. and Zurek, W. H. (Eds.) 1983, {\it Theory and
Measurement}, Princeton University Press, Princeton.

\end{thebibliography}

\newtheorem{theo}{Theorem}[section]

\newtheorem{definition}[theo]{Definition}

\newtheorem{lem}[theo]{Lemma}

\newtheorem{met}[theo]{Method}

\newtheorem{prop}[theo]{Proposition}

\newtheorem{coro}[theo]{Corollary}

\newtheorem{exam}[theo]{Example}

\newtheorem{rema}[theo]{Remark}{\hspace*{4mm}}

\newtheorem{example}[theo]{Example}

\newcommand{\proof}{\noindent {\em Proof:\/}{\hspace*{4mm}}}

\newcommand{\qed}{\hfill$\Box$}

\newcommand{\ninv}{\mathord{\sim}} 

\newtheorem{postulate}[theo]{Postulate}

\section*{Introduction}

In \cite{daCostadeRonde13}, Newton da Costa together with the author of this paper argued in favor of the possibility to consider quantum superpositions in terms of a paraconsistent approach. We claimed that, even though most interpretations of quantum mechanics (QM) attempt to escape contradictions, there are many hints that indicate it could be worth while to engage in a research of this kind. Recently, Arenhart and Krause \cite{ArenhartKrause14a, ArenhartKrause14b} have raised several arguments against this approach. More specifically, taking into account the square of opposition, they have argued that quantum superpositions are better understood in terms of {\it contrariety propositions} rather than in terms of {\it contradictory propositions}. In \cite{deRonde14a} we defended the Paraconsistent Approach to Quantum Superpositions (PAQS) and provided arguments in favor of its development. We showed that: {\it i) Arenhart and Krause placed their obstacles from a specific metaphysical stance, which we characterized in terms of what we call the Orthodox Line of Research (OLR).} And also, {\it ii) That this is not necessarily the only possible line, and that a different one, namely, a Constructive Metaphysical Line of Research (CMLR) provides a different perspective in which PAQS can be regarded as a valuable prospect.} Furthermore, we explained how, within the CMLR, the problems and obstacles raised by Arenhart and Krause disappear. More specifically, we argued that the OLR is based in two main principles: \\

\begin{enumerate}
\item {\bf Quantum to Classical Limit:}  The principle that one can find a  bridge between classical mechanics and QM; i.e., that the main notions of classical physics can be used in order to explain quantum theory.\\
\end{enumerate}

\begin{enumerate}
\item[2.] {\bf Classical Physical Representation:} The principle that one needs to presuppose the classical physical representation ---structured through the metaphysics of entities together with the mode of being of actuality--- in any interpretation of QM.\\
\end{enumerate}

\noindent In this context, regarding quantum superpositions, the Measurement Problem (MP) is one of the main questions imposed by the OLR. Given the fact that QM describes mathematically the state in terms of a superposition, the question is why do we observe a single result instead of a superposition of them? Although the MP accepts the fact that there is something very weird about quantum superpositions, leaving aside their problematic meaning, it focuses on the justification of the actualization process. 

Taking distance from the OLR, the CMLR is based on three main presuppositions already put forward in \cite[pp. 56-57]{deRonde11}.\\

\begin{enumerate}
\item {\bf Closed Representational Stance:} Each physical theory is closed under its own formal and conceptual structure and provides access to a specific set of phenomena. The theory also provides the constraints to consider, explain and understand physical phenomenon. The understanding of a phenomena is always {\it local} for it refers to the closed structure given by the physical theory from which observations are determined.\\

\item {\bf Formalism and Empirical Adequacy:} The formalism of QM is able to provide (outstanding) empirically adequate results. Empirical adequacy determines the success of a theory and not its commitment to a certain presupposed conception of the world. Thus, it seems to us that the problem is not to find a new formalism. On the contrary, the `road signs' point in the direction that {\it we must stay close to the orthodox quantum formalism}.\\

\item {\bf Constructive Stance:} To learn about what the formalism of QM is telling us about reality we might be in need of {\it creating new (non-classical) physical concepts}.\\
\end{enumerate}

\noindent Changing the metaphysical standpoint, the CMLR presents a different questioning which assumes right from the start the need of bringing into stage a different metaphysical scheme to the one assumed by the OLR ---based on the representation provided by classical physics. What is needed, according to this stance, is a radical inversion of orthodoxy and its problems. Regarding quantum superpositions, instead of considering the MP we should invert the questioning ---changing the perspective--- and concentrate in the analysis of what we have called: `the superposition problem'.  

In a more recent paper  \cite{ArenhartKrause14c}, Arenhart and Krause have continued their analysis arguing against the notion of potentiality and power presented in \cite{deRonde13b} and discussed in \cite{deRonde14a}. In the present paper we attempt to analyze the notions of {\it modality}, {\it potentiality}, {\it power} and {\it contradiction} in QM, and provide further arguments of why the PAQS is better suited, than the Contrariety Approach to Quantum Superpositions (CAQS) proposed by Arenhart and Krause, to face the interpretational questions that quantum technology is forcing us to consider. The paper is organized as follows. In section 1, we discuss the physical representation of quantum superpositions. Section 2 analyzes the meaning of modality in QM and puts forward our interpretation in terms of `ontological potentiality'. In section 3, we discuss the meaning of the notion of `power' as a real physical existent. In section 4, we analyze two different approaches to quantum superpositions, the PAQS and the CAQS. In section 5, we provide the conclusions of the paper.

\section{The Physical Representation of Quantum Superpositions}

In \cite{deRonde14a} we made it clear why we are interested ---through the CMLR--- in attacking the Superposition Problem (SP), which attempts to develop a physical representation of quantum superpositions, instead of discussing the famous Measurement Problem (MP) which ---following the OLR--- attempts to justify the actual non-contradictory realm of existence. The idea that quantum superpositions cannot be physically represented was stated already in 1930 by Paul Dirac in the first edition of his famous book: {\it The Principles of Quantum Mechanics}.  

\begin{quotation}
{\small``The nature of the relationships which the superposition principle requires to exist between the states of any system is of a kind that cannot be explained in terms of familiar physical concepts. One cannot in the classical sense picture a system being partly in each of two states and see the equivalence of this to the system being completely in some other state. There is an entirely new idea involved, to which one must get accustomed and in terms of which {\it one must proceed to build up an exact mathematical theory, without having any detailed classical picture}." \cite[p. 12]{Dirac74} (emphasis added)}\end{quotation}

\noindent Also Niels Bohr was eager to defend the classical physical representation of our world and set the limits of such representation in classical physics itself \cite{BokulichBokulich}. Bohr would set the problems of the present OLR by claiming explicitly that: \cite[p. 7]{WZ} ``[...] the unambiguous interpretation  of any measurement must be essentially framed in terms of classical physical theories, and we may say that in this sense the language of Newton and Maxwell will remain the language of physicists for all time.'' At the same time he closed any further conceptual development by arguing that ``it would be a misconception to believe that the difficulties of the atomic theory may be evaded by eventually replacing the concepts of classical physics by new conceptual forms.'' Even Erwin Schr\"odinger, who was one of the first to see the implications of {\bf the superposition principle} with his also famous ``cat experiment'', did not dare to think beyond the representation of classical physics \cite{Schr35}. 

Unfortunately, these ideas have sedimented in the present foundational research regarding QM. Indeed, the strategy of the OLR has been to presuppose the classical representation provided by classical Newtonian mechanics in terms of an {\it Actual State of Affairs}. The first problem is the so called ``basis problem'' which attempts to explain how nature ``chooses'' a single basis ---between the many possible ones--- when an experimental arrangement is determined in the laboratory ---this also relates to the problem of contextuality which we have analyzed in detail in \cite{deRondePHD}. The second problem is the already mentioned MP: given the fact that QM describes mathematically the state in terms of a superposition of states, the question is why do we observe a single result instead of a superposition of them? It should be clear that the MP presupposes the determination of a basis and is not related to contextuality nor the BP. Although the MP accepts the fact that there is something very weird about quantum superpositions, leaving aside their problematic physical meaning, it focuses on the justification of the actualization process. Taking as a standpoint the single outcome it asks: how do we get to a single measurement result from the quantum superposition?\footnote{The questioning is completely analogous to the one posed by the quantum to classical limit problem: how do we get from contextual weird QM into our safe classical physical description of the world?} The MP attempts to justify why, regardless of QM, we only observe actuality. The problem places the result in the origin, and what needs to be justified is the already known answer. 

Our interest, contrary to the OLR, focuses instead on what we have called the Superposition Problem (SP). According to it, in case one attempts to provide a realist account of QM, one should concentrate in finding a set of physical concepts which provide a physical representation of quantum superpositions. But in order to do so we need to go beyond the question regarding measurement outcomes. Before we can understand actualization we first need to explain what a quantum superposition {\it is} or {\it represents}. As we have argued elsewhere  \cite{deRonde13a}, there is no self evident path between the superposition and its outcome for there are multiple ways of understanding the {\it projection postulate} (see for a discussion \cite{RFD14a}).  

Our research is focused on the idea that quantum superpositions relate to something physically real that exists in Nature, and that in order to understand QM we need to provide a physical representation of such existent (see for a more detailed discussion \cite{deRonde15a}). But why do we think we have good reasons to believe that quantum superpositions exist? Mainly because quantum superpositions are one of the main sources used by present experimental physicists to develop the most outstanding technical developments and experiments of the last decades. Indeed, there are many characteristics and behaviors we have learnt about superpositions: we know about {\it their existence regardless of the effectuation of one of its terms}, as shown, for example, by the interference of different possibilities in {\it welcher-weg} type experiments \cite{ Nature11a, NaturePhy12}, {\it their reference to contradictory properties}, as in Schr\"{o}dinger cat states \cite{Nature07}, we also know about {\it their non-standard route to actuality}, as explicitly shown by the MKS theorem \cite{DFR06, RFD14a}, and we even know about {\it their non-classical interference with themselves and with other superpositions}, used today within the latest technical developments in quantum information processing \cite{Nature13}. In spite of the fact we still cannot say what a quantum superposition {\it is} or {\it represents}, we must admit that they seem ontologically robust. If the terms within a quantum superposition are considered as quantum possibilities (of being actualized) then we must also admit that such quantum possibilities {\it interact} ---according to the Schr\"odinger equation. It is also well known that one can produce interactions between multiple superpositions (entanglement) and then calculate the behavior of all terms as well as the ratio of all possible outcomes. It then becomes difficult not to believe that these terms that `interact', `evolve' and `can be predicted' according to the theory, are not real (in some way). 

Disregarding these known facts, most interpretations of QM do not consider quantum superpositions as related to physical reality. For example, the so called Copenhagen interpretation remains agnostic with respect to the mode of existence of properties {\it prior} to measurement. The same interpretation is endorsed by van Fraassen in his Copenhagen modal variant.\footnote{According to Van Fraassen \cite[p. 280]{VF91}: ``The interpretational question facing us is exactly: in general, which value attributions are true? The response to this question can be very conservative or very liberal. Both court later puzzles. I take it that the Copenhagen interpretation ---really, a roughly correlated set of attitudes expressed by members of the Copenhagen school, and not a precise interpretation--- introduced great conservatism in this respect. Copenhagen scientists appeared to doubt or deny that observables even have values, unless their state forces to say so. I shall accordingly refer to the following very cautious answer as the {\it Copenhagen variant} of the modal interpretation. It is the variant I prefer.''} Much more extreme is the instrumentalist perspective put forward by Fuchs and Peres \cite[p. 1]{FuchsPeres00} who claim that: ``[...] quantum theory does not describe physical reality. What it does is provide an algorithm for computing probabilities for the macroscopic events (``detector clicks") that are the consequences of experimental interventions.'' In Dieks' realistic modal version quantum superpositions are not considered as physical existents, only one of them is real (actual), namely, the one written as a single term, while all other superpositions of more than one term are considered as possible (in the classical sense). It seems then difficult to explain, through this interpretation, what is happening to the rest of non-actual terms which can be also predicted ---even though not with certainty. In a similar vein, the consistent histories interpretation developed by Griffiths and Omn\`es also argues against quantum superpositions \cite{Griffiths13}.\footnote{For a detailed analysis of the arguments provided by Dieks and Griffiths see: \cite{deRonde15a}.} Bohmian mechanics proposes the change the formalism and talk instead of a quantum field that governs the evolution of particles. 

One might also argue that some interpretations, although not explicitly, leave space to consider superpositions as existent in a potential, propensity, dispositional or latent realm. The Jauch and Piron School, Popper or Margenau's interpretations, are good examples of such proposals (see for discussion \cite{deRondePHD} and references therein). However,  within such interpretations the collapse is accepted and potentialities, propensities or dispositions are only defined in terms of `their becoming actual' ---mainly because, forced by the OLR, they have been only focused in providing an answer to the MP. In any case, such realms are not articulated nor physically represented beyond their meaning in terms of the actual realm. Only the many worlds interpretation goes as far as claiming that all terms in the superposition are real in actuality. However, the quite expensive metaphysical price to pay is to argue that there is a multiplicity of unobservable Worlds (branches) in which each one of the terms is actual. Thus, the superposition expresses the multiplicity of such classical actual Worlds.   

Instead of taking one of these two paths which force us either into the abandonment of representation and physical reality or to the exclusive account of physical representation in terms of an ASA, we have proposed through the CMLR to develop a new path which concentrates in developing radically new (non-classical) concepts.

\section{Modality and Ontological Potentiality in Quantum Mechanics}

QM has been related to modality since its origin, when Max Born interpreted Schr\"odinger's quantum wave function, $\Psi$, as a ``probability wave''. However, it was very clear from the very begining that the meaning of modality and probability in the context of QM was something completely new. As remarked by Heisenberg himself: 

\begin{quotation}
{\small ``[The] concept of the probability wave [in quantum mechanics] was something entirely new in theoretical physics since Newton. Probability in mathematics or in statistical mechanics means a statement about our degree of knowledge of the actual situation. In throwing dice we do not know the fine details of the motion of our hands which determine the fall of the dice and therefore we say that the probability for throwing a special number is just one in six. The probability wave function, however, meant more than that; it meant a tendency for something.'' \cite[p. 42]{Heis58}}\end{quotation}

\noindent Today, it is well known that quantum probability does not allow an interpretation in terms of ignorance \cite{Redei} ---even though many papers in the literature still use probability uncritically in this way. Instead, as we mentioned above, the quantum formalism seems to imply some kind of weird {\it interaction of possibilities} governed by the Schr\"odinger equation. 

So they say, we do not understand QM and trying to do so almost makes no sense since it is too difficult problem to be solved. If Einstein, Bohr, Heisenberg, Schr\"odinger and many of the most brilliant minds in the last century could not find an answer to this problem, maybe it is better to leave it aside. In line with these ideas, the problems put forward by the OLR have left behind the development of a new physical representation of QM and have instead concentrated their efforts in justifying our classical world of entities in the actual mode of existence. Only when leaving behind the OLR, one might be able to consider the possibility to provide a new non-classical physical representation of QM. Of course this implies reconsidering the meaning of existence itself and the abandonment of another presupposed dogma: existence and reality are represented by actuality either as an observation {\it hic et nunc} (empiricism) or as a mode of existence (classical realism).   

Following the CMLR, we believe a reasonable strategy would then be to start with what we know works perfectly well, namely, the orthodox formalism of QM and advance in the metaphysical principles which constitute our understanding of the theory. Escaping the metaphysics of actuality and starting from the formalism, a good candidate to develop a mode of existence is of course {\it quantum possibility}. In several papers, together with Domenech and Freytes, we have analyzed how to understand possibility in the context of the orthodox formalism of QM \cite{DFR06, DFR08a, DFR08b, DFR09}. From this investigation there are several conclusions which can be drawn. We started our analysis with a question regarding the contextual aspect of possibility. As it is well known, the Kochen-Specker (KS) theorem does not talk about probabilities, but rather about the constraints of the formalism to actual definite valued properties considered from multiple contexts (see for an extensive discussion regarding the meaning of contextuality \cite{deRonde15a}). What we found via the analysis of possible families of valuations is that a theorem which we called ---for obvious reasons--- the Modal KS (MKS) theorem can be derived which proves that quantum possibility, contrary to classical possibility, is also contextually constrained \cite{DFR06}. This means that, regardless of its use in the literature: {\it quantum possibility} is not {\it classical possibility}. In a recent paper, \cite{RFD14a} we have concentrated in the analysis of {\it actualization} within the orthodox frame and interpreted, following the structure, the logical realm of possibility in terms of potentiality.

Once we accept we have two distinct realms of existence: potentiality and actuality, we must be careful about the way in which we define {\it contradictions}. Certainly, contradictions cannot be defined in terms of truth valuations in the actual realm, simply because the physical notion that must be related to quantum superpositions must be, according to our research, an existent in the potential realm ---not in the actual one. The MKS theorem shows explicitly that a quantum wave function implies multiple incompatible valuations which can be interpreted as {\it potential contradictions}. Our analysis has always kept in mind the idea that contradictions ---by definition--- are never found in the actual realm. Our attempt is to turn things upside-down: we do not need to explain the actual via the potential but rather, we need to use the actual in order to develop the potential \cite[p. 148]{deRondePHD}. Leaving aside the paranoia against contradictions, the PAQS does the job of allowing a further formal development of a realm in which all terms of a superposition exist, regardless of actuality. In the sense just discussed the PAQS opens possibilities of development which have not yet been fully investigated. It should be also clear that we are not claiming that all terms in the superposition are actual ---as in the many worlds interpretations--- overpopulating existence with unobservable actualities. What we claim is that PAQS opens the door to consider all terms as existent in potentiality ---independently of actuality. We claim that just like we need all properties to characterize a physical object, all terms in the superposition are needed for a proper characterization of what exists according to QM. Contrary to Arenhart and Krause we do not agree that our proposal is subject of Priest's razor, the metaphysical principle according to which we should not populate the world with contradictions beyond necessity \cite{Priest87}. The PAQS does not overpopulate metaphysically the world with contradictions, rather it attempts to take into account what the quantum formalism and present experiments seem to be telling us about physical reality.\footnote{Regarding observation it is important to remark that such contradictory potentialities are observable just in the same way as actual properties can be observed in an object. Potentialities can be observed through actual effectuations in analogous fashion to physical objects ---we never observe all perspectives of an object simultaneously, instead, we observe at most a single set of actual properties.}

Modal interpretations are difficult to define within the literature.\footnote{As we have discussed in \cite{deRonde10} modal interpretations range from empiricist positions such as that of Van Fraassen \cite{VF91} to realist ones such as the one endorsed in different ways by Dieks \cite{Dieks89a}, Bub \cite{Bub97} and Bacciagaluppi \cite{Bacciagaluppi96}. There are even different strategies and ideas regarding what should be considered to be a coherent interpretation within this group.} We understand that modal interpretations have two main desiderata that must be fulfilled by any interpretation which deserves being part of the club. The first is to stay close to the standard formalism of QM, the second is to investigate the meaning of modality and existence within the orthodox formalism of the theory. The modal interpretation that we have proposed \cite{deRonde11} attempts to develop ---following these two desiderata and the CMLR--- a physical representation of the formalism based on two main notions: the notion of `ontological potentiality' and notion of `power'. The notion of ontological potentiality has been explicitly developed taking into account what we have learnt from the orthodox formalism about quantum possibility, taking potentiality to its limit and escaping the dogmatic ruling of actuality. Contrary to the teleological notion of potentiality used within many interpretations of QM our notion of ontological potentiality is not defined in terms of actuality \cite{Smets05}. Such perspective has determined not only the need to consider what we call a {\it Potential State of Affairs} ---in analogous fashion to the {\it Actual State of Affairs} considered within physical theories---, but also the distinction between {\it actual effectuations}, which is the effectuation of potentiality in the actual realm, and {\it potential effectuations} which is that which happens in the potential realm regardless of actuality \cite{deRonde13a, deRonde13b, RFD14a}. Actualization only discusses the actual effectuation of the potential, while potential effectuations remain in the potential realm evolving according to QM. The question we would like to discuss in the following section is: what is that which exists in the potential realm?

\section{Powers as Real Quantum Physical Existents}

Entities are composed by properties which exist in the actual mode of being. But what is that which exists in the ontological potential realm? We have argued that an interesting candidate to consider is the notion of {\it power}. Elsewhere \cite{deRonde11, deRonde13b}, we have put forward such an ontological interpretation of powers. In the following we summarize such ideas and provide an axiomatic characterization of QM in line with these concepts. 

{\it The mode of being of a power is potentiality}, not thought in terms of classical possibility (which relies on actuality) but rather as a mode of existence ---i.e., in terms of ontological potentiality. To possess the power of {\it raising my hand}, does not mean that in the future `I {\it will} raise my hand' {\it or} that in the future `I {\it will not} raise my hand'; what it means is that, here and now, I possess a power which exists in the mode of being of potentiality, {\it independently of what will happen in actuality}. Powers do not exist in the mode of being of actuality, they are not actual existents, they are undetermined potential existents. Powers, like classical properties, preexist to observation, unlike them, preexistence is not defined in the actual mode of being as an Actual State of Affairs (ASA), instead we have a {\it potential preexistence} of powers which determines a Potential State of Affairs (PSA). {\it Powers are indetermined.} Powers are a conceptual machinery which can allow us to compress experience into a picture of the world, just like entities such as particles, waves and fields, allow us to do so in classical physics. We cannot ``see" powers in the same way we see objects.\footnote{It is important to notice there is no difference in this point with the case of entities: we cannot ``see" entities ---not in the sense of having a complete access to them. We only see perspectives which are unified through the notion of object.} Powers are experienced in actuality through {\it elementary processess}. A power is sustained by a logic of actions which do not necessarily take place, it \emph{is} and \emph{is not}, {\it hic et nunc}. 

A basic question which we have posed to ourselves regards the ontological meaning of a {\it quantum superposition} \cite{deRonde11}. What does it mean to have a mathematical expression such as: $\alpha| \uparrow \rangle + \beta |\downarrow\rangle$, which allows us to predict precisely, according to the Born rule, experimental outcomes? Our theory of powers has been explicitly developed in order to try to answer this particular question. Given a superposition in a  particular basis, $\Sigma \ c_i | \alpha_i \rangle$, the powers are represented by the elements of the basis, $| \alpha_i \rangle$, while the coordinates in square modulus, $|c_i|^2$, are interpreted as the potentia of each respective power. {\it Powers can be superposed to different ---even contradictory--- powers.} We understand a quantum superposition as encoding a set of powers each of which possesses a definite {\it potentia}. This we call a {\it Quantum Situation} ($QS$). For example, the quantum situation represented by the superposition $\alpha | \uparrow \rangle + \beta |\downarrow\rangle$, combines the contradictory powers, $| \uparrow \rangle$ and $|\downarrow\rangle$, with their potentia, $|\alpha|^2$ and $|\beta|^2$, respectively. Contrary to the orthodox interpretation of the quantum state, we do not assume the metaphysical identity of the multiple mathematical representations given by different basis \cite{deRondeMassri14}. Each superposition is basis dependent and must be considered as a distinct quantum situation. For example, the superpositions $c_{x1} | \uparrow_{x} \rangle + c_{x2} |\downarrow_{x}\rangle$ and $c_{y1} | \uparrow_{y} \rangle + c_{y2} |\downarrow_{y}\rangle$, which are representations of the same $\Psi$ and can be derived from one another via a change in basis, are considered as two different and distinct quantum situations, $QS_{\Psi, B_x}$ and $QS_{\Psi, B_y}$. 

The logical structure of a superposition is such that a power and its opposite can exist at one and the same time, violating the principle of non-contradiction \cite{daCostadeRonde13}. Within the faculty of raising my hand, both powers (i.e., the power `I am {\it able to} raise my hand' and the power `I am {\it able not to} raise my hand') co-exist. A $QS$ is {\it compressed activity}, something which {\it is} and {\it is not} the case, {\it hic et nunc}. It cannot be thought in terms of identity but is expressed as a difference, as a {\it quantum}.

Our interpretation can be condensed in the following eight postulates. 

\begin{enumerate}

{\bf \item[I.] Hilbert Space:} QM is represented in a vector Hilbert space.\\

{\bf \item[II.] Potential State of Affairs (PSA):} A specific vector $\Psi$ with no given mathematical representation (basis) in Hilbert space represents a PSA; i. e., the definite existence of a multiplicity of powers, each one of them with a specific potentia.\\

{\bf \item[III.] Actual State of Affairs (ASA):} Given a PSA and the choice of a definite basis ${B, B', B'',...}$ (or equivalently a C.S.C.O.) a context is defined in which a set of powers, each one of them with a definite potentia, are univocally determined as related to a specific experimental arrangement (which in turn corresponds to a definite ASA). The context builds a bridge between the potential and the actual realms, between quantum powers and classical objects. The experimental arrangement (in the ASA) allows the powers (in the PSA) to express themselves in actuality through elementary processes which produce actual effectuations.\\

{\bf \item[IV.] Quantum Situations, Powers and Potentia:} Given a PSA, $\Psi$, and the context we call a quantum situation to any superposition of one or more than one power. In general given the basis $B= \{ | \alpha_i \rangle \}$ the quantum situation $QS_{\Psi, B}$ is represented by the following superposition of powers:
\begin{equation}
c_{1} | \alpha_{1} \rangle + c_{2} | \alpha_{2} \rangle + ... + c_{n} | \alpha_{n} \rangle
\end{equation}

\noindent We write the quantum situation of the PSA, $\Psi$, in the context $B$ in terms of the order pair given by the elements of the basis and the coordinates in square modulus of the PSA in that basis:
\begin{equation}
QS_{\Psi, B} = (| \alpha_{i} \rangle, |c_{i}|^2)
\end{equation}

\noindent The elements of the basis, $| \alpha_{i} \rangle$, are interpreted in terms of {\it powers}. The coordinates of the elements of the basis, $|c_{i}|^2$, are interpreted as the {\it potentia} of the power $| \alpha_{i} \rangle$, respectively. Given the PSA and the context, the quantum situation, $QS_{\Psi, B}$, is univocally determined in terms of a set of powers and their respective potentia. (Notice that in contradistinction with the notion of quantum state the definition of a quantum situation is basis dependent.)\\

{\bf \item[V.] Elementary Process:} In QM one can observe discrete shifts of energy (quantum postulate). These discrete shifts are interpreted in terms of {\it elementary processes} which produce actual effectuations. An elementary process is the path which undertakes a power from the potential realm to its actual effectuation. This path is governed by the {\it immanent cause} which allows the power to remain preexistent in the potential realm independently of its actual effectuation. Each power $| \alpha_{i} \rangle$ is univocally related to an elementary process represented by the projection operator $P_{\alpha_{i}} = | \alpha_{i} \rangle \langle \alpha_{i} |$.\\

{\bf \item[VI.] Actual Effectuation of Powers (Measurement):} Powers exist in the mode of being of ontological potentiality. An {\it actual effectuation} is the expression of a specific power in actuality. Different actual effectuations expose the different powers of a given $QS$. In order to learn about a specific PSA (constituted by a set of powers and their potentia) we must measure repeatedly the actual effectuations of each power exposed in the laboratory. (Notice that we consider a laboratory as constituted by the set of all possible experimental arrangements that can be related to the same $\Psi$.)\\ 

{\bf \item[VII.] Potentia (Born Rule):} A {\it potentia} is the strength of a power to exist in the potential realm and to express itself in the actual realm. Given a PSA, the potentia is represented via the Born rule. The potentia $p_{i}$ of the power $| \alpha_{i} \rangle$ in the specific PSA, $\Psi$, is given by:
\begin{equation}
Potentia \ (| \alpha_{i} \rangle, \Psi) = \langle \Psi | P_{\alpha_{i}} | \Psi \rangle = Tr[P_{ \Psi} P_{\alpha_{i}}]
\end{equation}

\noindent In order to learn about a $QS$ we must observe not only its powers (which are exposed in actuality through actual effectuations) but we must also measure the potentia of each respective power. In order to measure the potentia of each power we need to expose the $QS$ statistically through a repeated series of observations. The potentia, given by the Born rule, coincides with the probability frequency of repeated measurements when the number of observations goes to infinity.\\

{\bf \item[VIII.]  Potential Effectuation of Powers (Schr\"odinger Evolution):} Given a PSA, $\Psi$, powers and potentia evolve deterministically, independently of actual effectuations, producing {\it potential effectuations} according to the following unitary transformation:
\begin{equation}
i \hbar \frac{d}{dt} | \Psi (t) \rangle = H | \Psi (t) \rangle
\end{equation}
\noindent While {\it potential effectuations} evolve according to the Schr\"odinger equation, {\it actual effectuations} are particular expressions of each power (that constitutes the PSA, $\Psi$) in the actual realm. The ratio of such expressions in actuality is determined by the potentia of each power.\\ 
\end{enumerate}

\noindent According to our interpretation just like classical physics talks about entities composed by properties that preexist in the actual realm, QM talks about powers that preexist in in the (ontological) potential realm, independently of the specific actual context of inquiry. This interpretational move allows us to define powers independently of the context regaining an objective picture of physical reality independent of measurements and subjective choices. The price we are willing to pay is the abandonment of a metaphysical equation that has been presupposed in the analysis of the interpretation of QM: `actuality = reality'. In the following section, talking into account a typical quantum experience, we discuss in what sense powers are to be considered in terms of `contradiction' or `contrariety'.

\section{Contradiction and Contrariety in Quantum Superpositions}

Arenhart and Krause have called the attention to the understanding of contradiction via the Square of Opposition. 

\begin{quotation}
{\small ``States in quantum mechanics such as the one describing the famous Schr\"odinger cat ---which is in a superposition between the states `the cat is dead' and `the cat is alive'--- present a challenge for our understanding which may be approached via the conceptual tools provided by the square. According to some interpretations, such states represent contradictory properties of a system (for one such interpretation see, for instance, da Costa and de Ronde [6]). On the other hand, we have advanced the thesis that states such as `the cat is dead' and `the cat is alive' are contrary rather than contradictory (see Arenhart and Krause [1], [2]).'' \cite[p. 2]{ArenhartKrause14c}}\end{quotation}
 
\noindent Within their CAQS, Arenhart and Krause have argued in \cite{ArenhartKrause14c} against the concept of potentiality and its relation to contradiction concluding ``that contrariety is still a more adequate way to understand superpositions.'' Elsewhere, together with Domenech and Freytes, we have analyzed via the Square of Opposition the meaning of quantum possibility. We argued that the notion of possibility needs to be discussed in terms of the formal structure of the theory itself and that, in such case, one should not study the Classical Square of Opposition but rather an Orthomodular Square of Opposition. In \cite{FRD12} we developed such a structure and in \cite{RFD14b} we provided an interpretation of the Orthomodular Square of Opposition in terms of the notion of potentiality. Furthermore, according to the author of this paper, the development should also consider the analysis of the hexagon, paraconsistent negation and modalities provided by B\'eziau in  \cite{Beziau12, Beziau14}. In this section we argue that Arenhart and Krause have misinterpreted our notion of `potentiality' and `power' and explain why the PAQS is better suited to account for quantum superpositions than the CAQS. 

Let us begin our analysis recalling the traditional definitions of the famous square of opposition and the meaning of contradiction and contrariety.\\ 

\begin{enumerate}
\item[]{\bf Contradiction Propositions:} $\alpha$ and $\beta$ are {\it contradictory} when both cannot be true and both cannot be false.
\item[]{\bf Contrariety Propositions:} $\alpha$ and $\beta$ are {\it contrary} when both cannot be true, but both can be false.
\item[]{\bf Subcontrariety Propositions:} $\alpha$ and $\beta$ are {\it subcontraries} when both can be true, but both cannot be false.
\item[]{\bf Subaltern Propositions:} $\alpha$ is {\it subaltern} to proposition $\beta$ if the truth of $\beta$ implies the truth of $\alpha$.\\
\end{enumerate}

Discussing inadequacy of the notion of power, Arenhart and Krause provide the following analysis: 

\begin{quotation}
{\small ``First of all, a property, taken by itself as a power (a real entity not actual), is not affirmed nor denied of anything. To take properties such as `to have spin up in the x direction' and `to have spin down in the x direction' by themselves does not affirm nor deny anything. To say `to have spin up in the x direction' is not even a statement, it is analogous to speak `green' or `red hair'. To speak of a contradiction, it seems, one must have complete statements, where properties or relations are attributed to something. That is, one must have something like `spin up is measured in a given direction', or `Mary is red haired', otherwise there will be no occasion for truth and falsehood, and consequently, no occasion for a contradiction. So, the realm of the potential must be also a realm of attribution of properties to something if contradiction is to enter in it. However, this idea of attribution of properties seems to run counter the idea of a merely potential realm. On the other hand, the idea of a contradiction seems to require that we speak about truth and falsehood.'' \cite{ArenhartKrause14c}}\end{quotation}
 
The idea that potentiality determines a contradictory realm goes back to Aristotle himself who claimed that contradictions find themselves in potentiality. Of course, as remarked by Arenhart and Krause, the square of opposition is discussing about actual truth and falsehood. Thus, potentiality is not considered in terms of a mode of existence but rather as mere logical possibility. The interesting question is if our representation of quantum superpositions in terms of powers is compatible with the square. We believe it easy to see that such is the case provided special attention is given to the realms involved in the discussion. Furthermore, it is also easy to see that the CAQS is incompatible with QM due to its empirical inadequacy. Some remarks go in order. 
 
Firstly, we must stress the fact that a power is not ---as claimed by Arenhart and Krause [{\it Op. cit.}, p. 7]--- an entity. An entity exists in the mode of being of actuality and is represented by three main logical and ontological principles: {\bf the principle of existence}, {\bf the principle of non-contradiction} and {\bf the principle of identity} (see for discussion \cite{VerelstCoecke}). As discussed in the previous section, quite independently of such principles we have defined the notion of power in terms of {\bf the principle of indetermination}, {\bf the principle of superposition} and {\bf the principle of difference}. The adequacy or not of powers to interpret QM needs to be analyzed taking into account this specific scheme \cite{deRonde13b}. Instead of doing so, Arenhart and Krause have criticized a notion of power which they have not specified in rigorous terms.  

Secondly, truth and falsehood are related to actuality, since in the orthodox view this is the only exclusive realm considered as real. However, our notion of ontological potentiality is completely independent of actuality, and since powers are real objective existents it makes perfect sense to extend `truth' and `falsity' to this mode of being. It is the PSA which determines the specific set of powers which potentially preexist in a given quantum situation. Thus, in analogous fashion to the way an ASA determines the set of properties which are `true' and `false', a PSA determines a set of powers which are `true' and `false', namely, those powers which potentially preexist and can be exposed through the choice of different quantum situations (i.e., the multiplicity of possible contexts). Our redefinition of truth and falsehood with respect to potentiality escapes any subjective choice and regains an objective description of physical reality. In a given situation all the powers which determine possible actual effectuations compose a PSA. For example, a Stern Gerlach apparatus in a laboratory which can be placed in the $x$, $y$ or $z$ direction determines the existence of the powers: $ | \uparrow_x \rangle, | \downarrow_x \rangle, | \uparrow_y \rangle, | \downarrow_y \rangle, | \uparrow_z \rangle$ and $| \downarrow_z \rangle$ irrespectively of the actual choice of the particular context (i.e., the particular actual direction in which the Stern-Gerlach is placed). We can say that even though the PSA is defined independently of the context of inquiry, $QS$ are indeed {\it contextual existents}. 

Thirdly, let us investigate, provided we grant for the sake of the argument that powers do exist, which is the most suitable notion to account for two powers that can be actualized in a typical quantum experiment. Consider we have a Stern-Gerlach apparatus placed in the $x$ direction, if we have the following quantum superposition: $\alpha   | \uparrow_x \rangle +  \beta | \downarrow_x \rangle$, this means we have the power of having spin up in the $x$-direction, $| \uparrow_x \rangle$, with potentia $|\alpha|^2$ and the power of having spin down in the $x$-direction, $| \downarrow_x \rangle$, with potentia $|\beta|^2$. Both powers can become actual. Is it contradiction or contrariety the best notion suited to account for powers in this quantum experiment? Given this quantum situation, it is clear that both actualizations of the powers (elementary processes) $| \uparrow_x \rangle$ and $| \downarrow_x \rangle$ {\it cannot be simultaneously `true' in actuality}, since only one of them will become actual; it is also the case that both actualizations of the powers (elementary processes) $| \uparrow_x \rangle$ and $| \downarrow_x \rangle$ {\it cannot be simultaneously `false' in actuality}, since when we measure this quantum situation we know we will obtain either the elementary process `spin up in the $x$-direction', $| \uparrow_x \rangle\langle \uparrow_x |$, or the  elementary process `spin down in the $x$-direction', $| \downarrow_x \rangle\langle \downarrow_x |$. Now, if we consider the CAQS, contrary propositions are determined when {\it both cannot be true}, but {\it both can be false}. But this is not the case in QM, in particular, it is not the case for the example we have just considered. Given a measurement on the quantum superposition, $\alpha | \uparrow_x \rangle +   \beta | \downarrow_x \rangle$,  one of the two terms will become actual (true) while the other term will not be actual (false), which implies that {\it both cannot be false}. Thus, while the PAQS is able to describe what we know about what happens in a typical quantum measurement, the CAQS of Arenhart and Krause is not capable of fulfilling empirical adequacy. 

In the conclusion of their paper Arenhart and Krasue discuss what happens when the state is in a superposition. They argue that one possibility is to claim that ``when not in an eigenstate the system does not have any of the properties associated with the superposition.'' According to them: ``This option is compatible with the claim that states in a superposition are contraries: both fail to be the case.'' But as we have seen in the last section, given a superposition state such as $\alpha  |\uparrow_x \rangle +  \beta |\downarrow_x \rangle$, we know with certainty that one of the terms will become actual if measured. Thus, it makes no sense to claim that both will ``fail to be the case''. The CAQS, fails to provide the empirical adequacy needed to account for basic quantum experiments. A second possibility is to ``assume another interpretation [...] and hold that even in a superposition one of the associated properties hold, even if not in an eigenstate.'' According to Arenhart and Krasue: ``Following this second option, notice, the understanding of superpositions as contraries still hold: even when one of the properties in a superposition hold, the other must not be the case.'' However, if only one of the properties is {\it true} then it seems difficult to explain how a property that {\it is the case} can interact with a property that {\it is not the case}. As we know, the interaction of superpositions happens between all terms in the superposition, the possibilities contained in the superposition may interfere between each other. The question then raises: how can something that exists interact with something else which does not exist?\footnote{For a detailed analysis of the relation between quantum superpositions and physical reality see: \cite{deRonde15b}.}

\section{Final Remarks}

Although we agree with Arenhart and Krause regarding the fact that the formal approach that we provided in \cite{daCostadeRonde13} was not completely adequate to the idea discussed here, we must also remark that we never claimed that this was the final formal description of quantum superpositions but rather a very first step in such paraconsistent development. In this respect, we believe that this approach is still in need of further development.\footnote{A possible development in line with the interpretation presented in this paper will be analyzed in \cite{daCostadeRonde15}.} However, we must also remark that the approach provides a suitable answer to the existence of the multiple terms in a quantum superposition, something that is needed in order to make sense about present and future quantum experiments and technical developments. We believe that the possibilities it might open deserve not only attention but also criticism. We thank Arenhart and Krause for their careful and incisive analysis.  

\section*{Acknowledgements} 

We would like to thank an anonymous referee for useful comments and remarks. This work was partially supported by the following grants: FWO project G.0405.08 and FWO-research community W0.030.06. CONICET RES. 4541-12 (2013-2014).

\end{document}